\documentstyle[twoside,12pt]{article}
\begin{document}
\large
\bibliographystyle{plain}

\begin{titlepage}
\hfill\begin{tabular}{l}HEPHY-PUB 611/94\\ UWThPh-1994-44\\ December 1994
\end{tabular}\\[5.5cm]
\begin{center}
{\Large\bf SEMI-RELATIVISTIC HAMILTONIANS OF APPARENTLY NONRELATIVISTIC
FORM}\\
\vspace{1.5cm}
{\Large\bf Wolfgang LUCHA}\\[.5cm]
{\large Institut f\"ur Hochenergiephysik,\\
\"Osterreichische Akademie der Wissenschaften,\\
Nikolsdorfergasse 18, A-1050 Wien, Austria}\\[1cm]
{\Large\bf Franz F.~SCH\"OBERL}\\[.5cm]
{\large Institut f\"ur Theoretische Physik,\\
Universit\"at Wien,\\
Boltzmanngasse 5, A-1090 Wien, Austria}\\[1.5cm]
{\bf Abstract}
\end{center}
\normalsize

We construct effective Hamiltonians which despite their apparently
nonrelativistic form incorporate relativistic effects by involving parameters
which depend on the relevant momentum. For some potentials the corresponding
energy eigenvalues may be determined analytically. Applied to two-particle
bound states, it turns out that in this way a nonrelativistic treatment may
indeed be able to simulate relativistic effects. Within the framework of
hadron spectroscopy, this lucky circumstance may be an explanation for the
sometimes extremely good predictions of nonrelativistic potential models even
in relativistic regions.
\end{titlepage}

\section{Introduction}

The fundamental disadvantage inherent to any (semi-) relativistically
consistent description of some quantum-theoretic system is obviously brought
about by the nonlocality of the ``square-root'' operator of the
relativistically correct kinetic energy, $\sqrt{{\bf p}^2 + m^2}$,
entering necessarily in the Hamiltonian $H$ which governs the dynamics of the
system under consideration. In contrast to the nonrelativistic limit,
obtained from the expansion of the square root up to the lowest ${\bf
p}^2$-dependent order, $\sqrt{{\bf p}^2 + m^2} = m + {\bf
p}^2/(2\,m) + \dots$, the presence of the relativistic kinetic-energy
operator prevents, in general, a thoroughly analytic discussion; one is
forced to rely on some numerical solution of the problem.

This inconvenience may be circumvented---at least in principle---by
approximating a given semi-relativistic Hamiltonian $H$ (incorporating, by
definition, relativistic kinematics) by the corresponding ``effectively
semi-relativistic'' Hamiltonian, formulated and investigated according to the
lines proposed in the present work. These effective Hamiltonians are
characterized by their rigorous maintenance of the easier-to-handle
nonrelativistic kinematics while resembling the relativistic formalisms to
the utmost possible extent by replacing their intrinsic parameters by
effective ones which depend in a well-defined manner on the square of the
relevant momentum ${\bf p}$.

In order to be as concrete as possible, we choose to illustrate our route of
constructing and evaluating these effectively semi-relativistic Hamiltonians
for the particular case of bound states of two particles of spin zero. For
simplicity, let us assume that the two constituents of these bound states are
of equal mass $m$; the generalization to different masses is then
straightforward. In the framework of a semi-relativistic description all the
forces acting between these two particles may be derivable from some
coordinate-dependent interaction potential $V({\bf x})$. Consequently, the
semi-relativistic Hamiltonian describing this system in the
center-of-momentum frame of its constituents is given by
\begin{equation}
H = 2\sqrt{{\bf p}^2 + m^2} + V({\bf x})\ .
\label{eq:hsemi-rel}
\end{equation}

The equation of motion resulting from this type of Hamiltonian is usually
called ``spinless Salpeter equation.'' As it stands, it represents a
standard approximation to the Bethe--Salpeter formalism for bound states
within some relativistic quantum field theory. It may be derived from the
Bethe--Salpeter equation \cite{salpeter51}
\begin{enumerate}
\item by eliminating---in accordance with the spirit of an instantaneous
interaction---any dependence on timelike variables, which leads to the
so-called ``Salpeter equation'' \cite{salpeter52}, and
\item by neglecting any reference to the spin degrees of freedom of the two
involved bound-state constituents and restricting to solutions corresponding
exclusively to positive energy.
\end{enumerate}

The outline of this paper is as follows. We introduce, in
Sect.~\ref{sec:effsemrelham}, the effectively semi-relativistic Hamiltonians
corresponding to the really semi-relativistic Hamiltonians $H$ of
Eq.~(\ref{eq:hsemi-rel}) in their most general form. In
Sect.~\ref{sec:genstrat}, we derive, for the special case of power-law
potentials, some sort of ``master equation'' for that central quantity the
knowledge of which enables us to imitate the effects of relativistic
kinematics within a formally nonrelativistic framework, namely, the
expectation value of the square of the momentum ${\bf p}$. From the
consideration of the most important prototypes of interaction potentials in
Sect.~\ref{sec:effhamappl}, we are led to conclude, in
Sect.~\ref{sec:effhamconcl}, that our effective Hamiltonians represent indeed
a viable alternative to the original semi-relativistic Hamiltonians
(\ref{eq:hsemi-rel}).

\section{Effectively Semi-Relativistic Hamiltonians}\label{sec:effsemrelham}

The main idea of our way of constructing effectively semi-relativistic
Hamiltonians has already been sketched in Refs.~\cite{lucha91,lucha92}. (For
a brief account of this procedure see also Ref.~\cite{lucha94como}.)

The starting point of this construction is a trivial but nevertheless
fundamental inequality. This inequality relates the expectation values of
both the first and second powers of any Hermitian (or, to be more precise,
self-adjoint) but otherwise arbitrary operator ${\cal O} = {\cal O}^\dagger$,
taken with respect to (at this stage) arbitrary Hilbert-space vectors
$|\rangle$ (in the domain of ${\cal O}$) normalized to unity; it reads
$$
|\langle{\cal O}\rangle| \le \sqrt{\langle{\cal O}^2\rangle}\ .
\label{eq:ineqop}
$$
Application of the above inequality to the relativistic kinetic-energy
operator $\sqrt{{\bf p}^2 + m^2}$ yields
$$
\left\langle\sqrt{{\bf p}^2 + m^2}\right\rangle
\le \sqrt{\langle{\bf p}^2\rangle + m^2}\ .
$$
By employing this inequality, we obtain for an arbitrary expectation value
$\langle H\rangle$ of the semi-relativistic Hamiltonian $H$,
Eq.~(\ref{eq:hsemi-rel}),
\begin{eqnarray}
\langle H\rangle
&=& 2\left\langle\sqrt{{\bf p}^2 + m^2}\right\rangle + \langle V\rangle
\nonumber\\
&\le& 2\,\sqrt{\langle{\bf p}^2\rangle + m^2} + \langle V\rangle \nonumber\\
&=& 2\,\frac{\langle{\bf p}^2\rangle + m^2}
{\sqrt{\langle{\bf p}^2\rangle + m^2}} + \langle V\rangle \nonumber\\
&=& \left\langle 2\,\frac{{\bf p}^2 + m^2}
{\sqrt{\langle{\bf p}^2\rangle + m^2}} + V\right\rangle\ .
\label{eq:hamexpectval}
\end{eqnarray}

{}From now on we specify the Hilbert-space vectors in all expectation values to
be the eigenstates of our Hamiltonian $H$. In this case the expectation value
of $H$, $\langle H\rangle$, as appearing, e.~g., in (\ref{eq:hamexpectval}),
becomes the corresponding semi-relativistic energy eigenvalue $E$, i.~e.,
$$
E \equiv \langle H\rangle\ ,
$$
and the inequality (\ref{eq:hamexpectval}) tells us that this energy
eigenvalue is bounded from above by \cite{lucha91,lucha92}
$$
E \le \left\langle 2\,\frac{{\bf p}^2 + m^2}
{\sqrt{\langle{\bf p}^2\rangle + m^2}} + V\right\rangle\ .
$$
The operator within brackets on the right-hand side of this inequality may be
regarded as some ``effectively semi-relativistic'' Hamiltonian $H_{\rm eff}$
which possesses, quite formally, the structure of a nonrelativistic
Hamiltonian \cite{lucha91,lucha92},
\begin{eqnarray}
H_{\rm eff}
&\equiv& 2\,\frac{{\bf p}^2 + m^2}{\sqrt{\langle{\bf p}^2\rangle + m^2}} + V
\nonumber\\
&=& 2\,\hat m + \frac{{\bf p}^2}{\hat m} + V_{\rm eff}\ ,
\label{eq:heffective}
\end{eqnarray}
but involves, however, the effective mass \cite{lucha91,lucha92}
\begin{equation}
\hat m = \frac{1}{2}{\sqrt{\langle{\bf p}^2\rangle + m^2}}
\label{eq:meffective}
\end{equation}
and the effective nonrelativistic potential \cite{lucha91,lucha92}
\begin{eqnarray}
V_{\rm eff} &=& \frac{2\,m^2}{\sqrt{\langle{\bf p}^2\rangle + m^2}}
- \sqrt{\langle{\bf p}^2\rangle + m^2} + V \nonumber\\
&=& 2\,\hat m - \frac{\langle{\bf p}^2\rangle}{\hat m} + V\ .
\label{eq:poteffective}
\end{eqnarray}
The effective mass $\hat m$ as given by Eq.~(\ref{eq:meffective}) as well as
the constant, i.~e., coordinate-independent, term in the effective potential
$V_{\rm eff}$ of Eq.~(\ref{eq:poteffective}),
$$
2\,\hat m - \frac{\langle{\bf p}^2\rangle}{\hat m}\ ,
$$
quite obviously depend on the expectation value of the square of the momentum
${\bf p}$, that is, on $\langle{\bf p}^2\rangle$, and will therefore differ
for different energy eigenstates.

Motivated by our above considerations, we propose to approximate the true
energy eigenvalues $E$ of the semi-relativistic Hamiltonian $H$ of
Eq.~(\ref{eq:hsemi-rel}) by the corresponding ``effective'' energy
eigenvalues $E_{\rm eff}$, defined as the expectation values of some
effective Hamiltonian $\tilde H_{\rm eff}$ taken with respect to the
eigenstates $|\rangle_{\rm eff}$ of its own,
$$
E_{\rm eff} = \langle\tilde H_{\rm eff}\rangle_{\rm eff}\ ,
$$
where the effective Hamiltonian $\tilde H_{\rm eff}$, as far as its form is
concerned, is given by Eqs.~(\ref{eq:heffective}) to (\ref{eq:poteffective})
but is implicitly understood to involve the expectation values of ${\bf
p}^2$ with respect to the effective eigenstates $|\rangle_{\rm eff}$
(that is, $\langle{\bf p}^2\rangle_{\rm eff}$ in place of $\langle{\bf
p}^2\rangle$):
$$
\tilde H_{\rm eff} = 4\,\tilde m
+ \frac{{\bf p}^2 - \langle{\bf p}^2\rangle_{\rm eff}}{\tilde m} + V\ ,
$$
with
$$
\tilde m = \frac{1}{2}{\sqrt{\langle{\bf p}^2\rangle_{\rm eff} + m^2}}\ .
$$
Accordingly, the effective energy eigenvalues $E_{\rm eff}$ are given by a
rather simple formal expression, viz., by
\begin{equation}
E_{\rm eff} = 4\,\tilde m + \langle V\rangle_{\rm eff}\ .
\label{eq:effenergy}
\end{equation}

\section{General Strategy of Evaluation}\label{sec:genstrat}

We intend to elaborate our general prescription for the construction of
effectively semi-relativistic Hamiltonians $\tilde H_{\rm eff}$ in more
detail for the particular case of power-law potentials depending only on the
radial coordinate $r \equiv |{\bf x}|$, i.~e., for potentials of the form
$$
V(r) = a\,r^n
$$
with some constant $a$. The reason for this restriction is twofold:
\begin{enumerate}
\item On the one hand, for power-law potentials the most general virial
theorem \cite{lucha89,lucha90mpla} in its nonrelativistic form
\cite{lucha91,lucha92} appropriate for the present case,
$$
\left\langle\frac{{\bf p}^2}{\tilde m}\right\rangle_{\rm eff}
= \frac{1}{2}\left\langle r\,\frac{dV(r)}{dr}\right\rangle_{\rm eff}\ ,
\label{eq:virial}
$$
enables us to replace the expectation value of the potential in
(\ref{eq:effenergy}) immediately by a well-defined function of the
expectation value of the squared momentum:
$$
a\,\langle r^n\rangle_{\rm eff}
= \frac{2}{n}\,\frac{\langle{\bf p}^2\rangle_{\rm eff}}{\tilde m}\ .
$$
This implies for the effective energy eigenvalues
\begin{equation}
E_{\rm eff} = 4\,\tilde m + \frac{2}{n}\,
\frac{\langle{\bf p}^2\rangle_{\rm eff}}{\tilde m}\ .
\label{eq:eeff}
\end{equation}
\item On the other hand, we may take advantage of the fact that for power-law
potentials it is possible to pass, without change of the fundamental
commutation relations between coordinate variables and their canonically
conjugated momenta, from the dimensional phase-space variables employed at
present to new, dimensionless phase-space variables and to rewrite the
Hamiltonian in form of a Hamiltonian which involves only these dimensionless
phase-space variables \cite{lucha91}. The eigenvalues $\epsilon$ of this
dimensionless Hamiltonian are, of course, also dimensionless \cite{lucha91}.
Applying this procedure, we find for the effective energy eigenvalues
\begin{eqnarray*}
E_{\rm eff} - 4\,\tilde m
+ \frac{\langle {\bf p}^2\rangle_{\rm eff}}{\tilde m}
&=& \left\langle\frac{{\bf p}^2}{\tilde m} + a\,r^n\right\rangle_{\rm eff} \\
&=& \left(\frac{a^2}{{\tilde m}^n}\right)^{1/(2+n)} \epsilon\ .
\end{eqnarray*}
\end{enumerate}
Combining both of the above expressions for $E_{\rm eff}$, we obtain a
relation which allows us to determine $\langle{\bf p}^2\rangle_{\rm eff}$
unambiguously in terms of the dimensionless energy eigenvalues $\epsilon$:
\begin{equation}
\langle{\bf p}^2\rangle_{\rm eff}^{2+n}
= \frac{1}{4}\left(\frac{n}{2+n}\right)^{2+n}a^2\,\epsilon^{2+n}
\left(\langle{\bf p}^2\rangle_{\rm eff} + m^2\right)\ .
\label{eq:master}
\end{equation}
For a given power $n$, this equation may be solved for $\langle{\bf
p}^2\rangle_{\rm eff}$. Insertion of the resulting expression into
Eq.~(\ref{eq:eeff}) then yields the corresponding eigenvalue $E_{\rm eff}$ of
the effectively semi-relativistic Hamiltonian $\tilde H_{\rm eff}$.

\section{Applications}\label{sec:effhamappl}

We would like to investigate the capabilities of the effective treatment
proposed in the previous sections by discussing some of its implications for
some familiar prototypes of interaction potentials, namely, for the
harmonic-oscillator, Coulomb, linear, and funnel potential. To this end we
compare for the lowest-lying energy eigenstates (which we will label
according to the usual spectroscopic notation) the energy eigenvalues
$E_{\rm eff}$ resulting from our effective description with the respective
energy eigenvalues $E_{\rm NR}$ obtained within the corresponding and by now
rather standard nonrelativistic approach \cite{lucha91,lucha92}.

Occasionally, it will prove to be favourable to inspect, in particular, the
ultrarelativistic limit of the developed effective formalism, defined by
vanishing mass $m$ of the bound-state constituents, i.~e., by $m = 0$.

The relevant parameter space of our effective Hamiltonian $\tilde H_{\rm
eff}$ with a power-law potential of the form $V(r) = a\,r^n$ is spanned by
the mass $m$ of the bound-state constituents and the coupling strength $a$ of
the potential. Quite obviously, the crucial question within this context is:
For a given level of excitation of the bound system and a given coupling
strength $a$, for which range of the mass $m$ does our effective treatment
represent indeed a better approximation to the correct semi-relativistic
description of the quantum system under consideration than the much more
simple-minded nonrelativistic approach?

Consequently, we compare in the following, for the above-mentioned prototype
potentials, the difference of the effective energy eigenvalues $E_{\rm eff}$
and the semi-relativistic energy eigenvalues $E$ with the difference of the
nonrelativistic energy eigenvalues $E_{\rm NR}$ and the semi-relativistic
energy eigenvalues $E$; in other words, we consider the ratio
\begin{equation}
R := \frac{E_{\rm eff} - E}{E_{\rm NR} - E}\ .
\label{eq:rodoel}
\end{equation}
As long as (the modulus of) this ratio $R$ is less than one, the errors of
the energy eigenvalues induced by our effective treatment are definitely
smaller than those brought about by the nonrelativistic approach.

The spectra of both nonrelativistic and effective energy eigenvalues of
harmonic-oscillator and Coulomb potential may be investigated on entirely
algebraic grounds. More sophisticated potentials as well as the
semi-relativistic spectra of harmonic-oscillator and Coulomb potential,
however, have to be handled with the help of numerical methods:
\begin{itemize}
\item The numerical results for all Schr\"odinger-like situations, that is,
for both the nonrelativistic and effective approaches to linear and funnel
potential as well as for the semi-relativistic approach to the
harmonic-oscillator potential---in which case the semi-relativistic
Hamiltonian may be transformed to the one of a nonrelativistic
Schr\"odinger-type problem---, have been computed in an iterative way with
the help of the numerical scheme developed in Ref.~\cite{falk85}.
\item The numerical results for the semi-relativistic treatment of linear and
funnel potential have been obtained by a procedure similar to the so-called
``method of orthogonal collocation'' \cite{orthcoll}. This method
approximates the action of the square-root operator $\sqrt{{\bf p}^2 + m^2}$
of the relativistic kinetic energy on some suitably chosen (truncated) set of
basis states by a well-defined (finite) matrix representation.
\end{itemize}

For obvious reasons, we do not attempt to fit the predicted effective energy
eigenvalues to some experimentally observed particle spectrum. Nevertheless,
for our numerical discussion we employ parameter values which indicate, at
least, the physically reasonable orders of magnitude. We increase the mass
$m$ of the bound-state constituents gradually from zero to $m = 1.8$ GeV,
which corresponds (roughly) to the typical mass of the constituent c quark,
while keeping the coupling constants in the considered potentials fixed at
some typical values suggested by various attempts of phenomenological
descriptions of hadrons as bound states of quarks by (nonrelativistic)
potential models \cite{lucha91,lucha92}.

For dimensional reasons, in the case of the Coulomb potential any kind of
energy eigenvalue must be necessarily proportional to the mass $m$ of the
bound-state constituents, which renders the energy-difference ratio $R$ given
by Eq.~(\ref{eq:rodoel}) independent of $m$. For the harmonic-oscillator,
linear, and funnel potentials, our numerical investigations result in the
following findings for the dependence of $R$ on the mass $m$ and on the level
of excitation:
\begin{enumerate}
\item There is a certain critical value of the mass $m$ of the bound-state
constituents, which depends, of course, on the considered level of excitation
and on the particular value of the coupling strength $a$ in the potential.
For particle masses $m$ smaller than this boundary mass, the ratio $R$
defined by Eq.~(\ref{eq:rodoel}) stays between 0 and 1. This means that below
this specific boundary mass the effective energy eigenvalues $E_{\rm eff}$
are, at least, closer to the (exact) semi-relativistic ones $E$ than their
nonrelativistic competitors $E_{\rm NR}$. Furthermore, the ratio $R$
decreases with decreasing mass $m$ of the bound-state constituents.
Consequently, a diminution of this mass $m$ certainly improves the quality of
the approximation induced by the effective formalism compared to the
nonrelativistic approach.
\item There appears to exist a general trend of the decrease of the ratio $R$
defined by Eq.~(\ref{eq:rodoel}) for successively higher levels of
excitation. As a consequence of this, the critical mass becomes the larger
the higher the excitation of the bound system under consideration is. Quite
obviously, this effect increases the range of applicability of our effective
formalism for higher levels of excitation.
\end{enumerate}

An important feature of the experimentally measured mass spectra of
hadrons---which may serve to provide a decisive criterion regarding the
usefulness of our effective treatment for a meaningful description of
hadrons---is the empirically well-established linearity of the Regge
trajectories: both mesons and baryons may be grouped to form sets of
particles which populate (approximately) linear Regge trajectories; the
different members of these sets are related by the fact that, apart from a
constant shift, the squares of their masses, i.~e., of the energy eigenvalues
of the corresponding bound states of quarks in their center-of-momentum
frame, are proportional to the relative orbital angular momentum $\ell$ of
the bound-state constituents or, equivalently, the spin of the composite
particles, with almost one and the same constant of proportionality, the
so-called Regge slope
$$
\beta \simeq 1.2 \mbox{ GeV}^2\ ,
$$
for all Regge trajectories \cite{pdg94}. We indicate these relationships by
$$
E^2(\ell) = \beta\,\ell + \mbox{const.}
$$
In general, the theoretical dependence of the energy eigenvalues $E$ on the
angular momentum $\ell$ will turn out to be described by some rather
complicated function of $\ell$. For this reason we only take a quick glance
on the asymptotic behaviour of the predicted energy eigenvalues $E(\ell)$ for
large values of the angular momentum $\ell$, symbolically denoted by the
limit $\ell \to \infty$. There we may expect to observe a simple power-law
rise of the calculated squares of energy eigenvalues $E^2(\ell)$ for
increasing values of $\ell$.

\subsection{Harmonic oscillator}

For the harmonic-oscillator potential $V(r) = a\,r^2$, that is, for $n = 2$,
Eq.~(\ref{eq:master}) reduces to a quartic equation for the expectation value
$\langle{\bf p}^2\rangle_{\rm eff}$. Inserting the well-known expression
\cite{lucha91} for the dimensionless energy eigenvalues $\epsilon$ of the
three-dimensional harmonic oscillator,
$$
\epsilon = 2\,N\ ,
$$
where the total quantum number $N$ is given in terms of the radial and
orbital angular-momentum quantum numbers $n_{\rm r}$ and $\ell$,
respectively, by
$$
N = 2\,n_{\rm r} + \ell + \frac{3}{2}\ ,
\quad n_{\rm r} = 0,1,2,\dots, \quad \ell = 0,1,2,\dots,
$$
it is a simple task to write down the analytic solution of this quartic
equation for $\langle{\bf p}^2\rangle_{\rm eff}$ in terms of the potential
parameter $a$, the mass $m$ of the bound-state constituents, and the above
total quantum number $N$. According to our above prescription, the effective
energy eigenvalue is then given by inserting this result into
Eq.~(\ref{eq:eeff}). In the ultrarelativistic limit this effective energy
eigenvalue takes a particularly simple form: for $m = 0$ one finds
$$
\langle{\bf p}^2\rangle_{\rm eff} = \left(\frac{a}{2}\right)^{2/3} N^{4/3}
$$
and
\begin{eqnarray*}
E_{\rm eff} &=& 4\sqrt{\langle{\bf p}^2\rangle_{\rm eff}} \\
&=& 2\,(4\,a)^{1/3}N^{2/3}\ .
\end{eqnarray*}

Table \ref{tab:ratio-harmosci} compares the nonrelativistic and effectively
semi-relativistic approaches for the harmonic oscillator, exemplifying
thereby the above findings 1 and 2.
{\normalsize
\begin{table}[hbt]
\begin{center}
\caption[]{Ratio $R$ of the differences between effective and
semi-relativistic and between nonrelativistic and semi-relativistic energy
eigenvalues, defined in Eq.~(\ref{eq:rodoel}), for the three lowest-lying
energy eigenstates (denoted by 1S, 1P, and 2S) of the harmonic-oscillator
potential $V(r) = a\,r^2$, with $a = 0.5 \mbox{ GeV}^3$ and increasing mass
$m$ of the bound-state constituents.}\label{tab:ratio-harmosci}
\vspace{0.5cm}
\begin{tabular}{|c|c|c|c|}
\hline\hline
&\multicolumn{3}{c|}{}\\[-1ex]
&\multicolumn{3}{c|}{State}\\
&\multicolumn{3}{c|}{}\\[-1.5ex]
\cline{2-4}
$\begin{array}{c}\mbox{$\qquad m$ [GeV]$\qquad$}\\[1ex]{}\end{array}$
&1S&1P&2S\\[-0.5ex]
\hline
&&&\\[-1.1ex]
$\qquad0.250\qquad$&$\qquad0.192\qquad$&$\qquad0.120\qquad$&
$\qquad0.122\qquad$\\
$\qquad0.336\qquad$&$\qquad0.250\qquad$&$\qquad0.156\qquad$&
$\qquad0.156\qquad$\\
$\qquad0.500\qquad$&$\qquad0.358\qquad$&$\qquad0.228\qquad$&
$\qquad0.218\qquad$\\
$\qquad0.750\qquad$&$\qquad0.515\qquad$&$\qquad0.339\qquad$&
$\qquad0.311\qquad$\\
$\qquad1.000\qquad$&$\qquad0.665\qquad$&$\qquad0.451\qquad$&
$\qquad0.401\qquad$\\
$\qquad1.800\qquad$&$\qquad1.147\qquad$&$\qquad0.812\qquad$&
$\qquad0.673\qquad$\\[1.3ex]
\hline\hline
\end{tabular}
\end{center}
\end{table}}

Furthermore, it is no problem to determine immediately the large-$\ell$
behaviour of the theoretical energy eigenvalues. In the ultrarelativistic
case, because of $N \propto \ell$ for large $\ell$, the effective energy
eigenvalues $E_{\rm eff}$ behave, according to their above-mentioned
explicit general form, like
$$
E_{\rm eff}^2(\ell) \propto \ell^{4/3}\ .
$$
In very clear contrast to that, the large-$\ell$ asymptotic behaviour of the
corresponding nonrelativistic energy eigenvalues $E_{\rm NR}$ is given by
\cite{lucha91,lucha92}
$$
E_{\rm NR} = 2\sqrt{\frac{a}{m}}\,\ell + \mbox{const.}\ ,
$$
which implies
$$
E_{\rm NR}^2(\ell) \propto \ell^2\ .
$$
Without really great surprise, we arrive at the satisfactory conclusion that
for the harmonic-oscillator potential (at least the ultrarelativistic limit
of) the effective treatment comes closer to the observed linearity of the
Regge trajectories than the nonrelativistic approach.

\subsection{Coulomb potential}

For the Coulomb potential $V(r) = - \kappa/r$, that is, for $n = - 1$,
Eq.~(\ref{eq:master}) reduces to a linear equation for the expectation value
$\langle{\bf p}^2\rangle_{\rm eff}$. Inserting the well-known expression
\cite{lucha91} for the dimensionless energy eigenvalues $\epsilon$ of the
nonrelativistic Coulomb problem,
$$
\epsilon = - \frac{1}{(2\,N)^2}\ ,
$$
where the total quantum number $N$ is given in terms of the radial and
orbital angular-momentum quantum numbers $n_{\rm r}$ and $\ell$,
respectively, by
$$
N = n_{\rm r} + \ell + 1\ ,
\quad n_{\rm r} = 0,1,2,\dots, \quad \ell = 0,1,2,\dots,
$$
we obtain from this linear equation for $\langle{\bf p}^2\rangle_{\rm eff}$
$$
\langle{\bf p}^2\rangle_{\rm eff}
= \frac{\kappa^2\,m^2}{16\,N^2 - \kappa^2}\ ,
$$
and, after inserting this expression into Eq.~(\ref{eq:eeff}), for the
effective energy eigenvalues
$$
E_{\rm eff}
= \frac{m}{N}\,\frac{8\,N^2 - \kappa^2}{\sqrt{16\,N^2 - \kappa^2}}\ .
$$

In the ultrarelativistic limit $m = 0$ all of these energy eigenvalues
vanish. For the Coulomb problem, because of the lack of any sort of
dimensional parameter inherent to the theory in the case $m = 0$, this kind
of degeneracy must take place already for dimensional reasons. It may be
understood completely by application of the most general, that is,
relativistic, so-called ``master'' virial theorem \cite{lucha89,lucha90mpla}
derived by the present authors.

Picking up the question of the large-$\ell$ behaviour of the theoretical
energy eigenvalues again, we find from the reported explicit expression that
in the limit $\ell \to \infty$ the effective energy eigenvalues
$E_{\rm eff}$ will not depend on the orbital angular momentum $\ell$ at all:
$$
E_{\rm eff}^2(\ell) \propto \ell^0\ .
$$
In the nonrelativistic case, on the other hand, the energy eigenvalues
$E_{\rm NR}$ behave asymptotically like \cite{lucha91,lucha92}
$$
E_{\rm NR} = - \frac{m\,\kappa^2}{4\,\ell^2} + \mbox{const.}\ ,
$$
which implies the asymptotic independence of also the nonrelativistic energy
eigenvalues $E_{\rm NR}$ of the orbital angular momentum $\ell$:
$$
E_{\rm NR}^2(\ell) \propto \ell^0\ .
$$

\subsection{Variational method}\label{subsec:varmeth}

In general, it will not be possible to find some analytic expressions for the
effective energy eigenvalues $E_{\rm eff}$. However, in order to obtain an
approximation to the spectrum of energy eigenvalues to be expected or to get,
at least, some idea of it one may adopt the variational method described in
the following.

This standard variational method proceeds along the steps of the following,
extremely simple recipe \cite{lucha92,flamm82}:
\begin{enumerate}
\item Choose a suitable set of trial states $\{|\lambda\rangle\}$. The
different members of this set $\{|\lambda\rangle\}$ are distinguished from
each other by some sort of variational parameter $\lambda$.
\item Compute the set of expectation values of the Hamiltonian under
consideration, $H$, with respect to these trial states $|\lambda\rangle$ in
order to obtain
$$
E(\lambda) \equiv \langle\lambda|H|\lambda\rangle\ .
$$
\item Determine, from the first derivative with respect to $\lambda$, that
value $\lambda_{\rm min}$ of the variational parameter $\lambda$ which
minimizes the resulting, $\lambda$-dependent expression $E(\lambda)$.
\item Compute $E(\lambda)$ at the point of the minimum $\lambda_{\rm min}$ to
find in this way the minimal expectation value $E(\lambda_{\rm min})$ of the
Hamiltonian $H$ in the Hilbert-space subsector of the chosen trial states
$|\lambda\rangle$.
\end{enumerate}
This minimum $E(\lambda_{\rm min})$ provides, of course, only an upper
bound\footnote{\normalsize\ The accuracy of this method is discussed in
Ref.~\cite{schoeberl82}.} to the proper energy eigenvalue $E$ of the
Hamiltonian $H$:
$$
E \le E(\lambda_{\rm min})\ .
$$

Application of this straightforward variational procedure to one of our
effectively semi-relativistic Hamiltonians $\tilde H_{\rm eff}$ leads to
$E_{\rm eff}(\lambda_{\rm min})$, which, according to its derivation,
represents at least an upper bound to the corresponding effective energy
eigenvalue $E_{\rm eff}$.

Note that, as far as the above variational procedure is concerned, the
expectation value $\langle{\bf p}^2\rangle_{\rm eff}$ entering in the
effective Hamiltonian has to be regarded as a constant. Consequently, it has
not to be taken into account in the course of minimization of the energy
expression $E(\lambda)$ by varying the characteristic parameter $\lambda$.
Rather, in the framework of this variational technique, it has to be equated
to the expectation value of ${\bf p}^2$ taken with respect to precisely
that trial state $|\lambda_{\rm min}\rangle$ which is characterized by just
the minimizing value $\lambda_{\rm min}$ of the variational parameter
$\lambda$, that is, to $\langle\lambda_{\rm min}|{\bf
p}^2|\lambda_{\rm min}\rangle$.

For the present investigation we adopt the simplest conceivable set of trial
states $|\lambda\rangle$, namely, the ones the coordinate-space
representation $\psi({\bf x})$ of which is given, for a vanishing radial
quantum number $n_{\rm r}$, by the Gaussian trial functions (w.~l.~o.~g.,
$\lambda > 0$)
$$
\psi_{\ell m}(r,\theta,\phi)
= \sqrt{\frac{2\,\lambda^{2\,\ell + 3}}
{\Gamma\left(\ell + \mbox{$\frac{3}{2}$}\right)}}\,r^\ell\,
\exp\left(- \frac{\lambda^2\,r^2}{2}\right)
{\cal Y}_{\ell m}(\theta,\phi)\ ,
$$
where ${\cal Y}_{\ell m}$ denote the spherical harmonics for angular momentum
$\ell$ and projection $m$, and the normalization factor of these trial
functions makes use of the so-called gamma function \cite{abramow}
$$
\Gamma(z) \equiv \int\limits_0^\infty dt\,t^{z - 1}\,\exp(- t)\ .
$$
For this particular set of trial functions we obtain for the expectation
values of the square ${\bf p}^2$ of the momentum ${\bf p}$ and of the
$n$-th power $r^n$ of the radial coordinate $r$, respectively, with respect
to the trial states $|\lambda\rangle$:
$$
\langle\lambda|{\bf p}^2|\lambda\rangle
= \left(\ell + \mbox{$\frac{3}{2}$}\right)\lambda^2
$$
and
$$
\langle\lambda|r^n|\lambda\rangle
= \frac{\Gamma\left(\ell + \mbox{$\frac{3 + n}{2}$}\right)}
{\Gamma\left(\ell + \mbox{$\frac{3}{2}$}\right)}\,
\frac{1}{\lambda^n}\ .
$$

\subsection{Linear potential}

For the linear potential $V(r) = a\,r$, that is, for $n = 1$,
Eq.~(\ref{eq:master}) reduces to a cubic equation for the expectation value
$\langle{\bf p}^2\rangle_{\rm eff}$ which, of course, may be solved
analytically. Unfortunately, for the linear potential the dimensionless
energy eigenvalues $\epsilon$ are only known \cite{lucha91} for the case of
vanishing orbital angular momentum $\ell$, i.~e., only for $\ell = 0$. In
this case they are given by the negative zeros of the Airy function
\cite{abramow}. In any case, that is, for arbitrary values of $\ell$, the
effective energy eigenvalues $E_{\rm eff}$ may be found by employing some
numerical procedure.

However, before performing a numerical computation of the energy eigenvalues
$E_{\rm eff}$ of the effective Hamiltonian $\tilde H_{\rm eff}$ with linear
potential, we apply the simple variational technique introduced in the
preceding subsection. For this Hamiltonian the value of the variational
parameter $\lambda$ which minimizes the relevant expectation value
$\langle\lambda|\tilde H_{\rm eff}|\lambda\rangle$, that is,
$\lambda_{\rm min}$, is implicitly given by
$$
\lambda_{\rm min}^3 = \frac{a}{2}\,\frac{\Gamma(\ell + 2)}
{\Gamma\left(\ell + \mbox{$\frac{5}{2}$}\right)}\,\tilde m\ .
$$
Recalling the definition of $\tilde m$ as given in
Sect.~\ref{sec:effsemrelham}, we obtain from this expression a cubic equation
for $\langle\lambda_{\rm min}|{\bf p}^2|\lambda_{\rm min}\rangle$,
$$
\langle\lambda_{\rm min}|{\bf p}^2|\lambda_{\rm min}\rangle^3
= \frac{a^2}{16}\left(\ell + \mbox{$\frac{3}{2}$}\right)
\left(\frac{\Gamma(\ell + 2)}
{\Gamma\left(\ell + \mbox{$\frac{3}{2}$}\right)}\right)^2\left(
\langle\lambda_{\rm min}|{\bf p}^2|\lambda_{\rm min}\rangle + m^2\right)\ ,
$$
the analytic solution of which may be written down quickly. Insertion of this
result into Eq.~(\ref{eq:eeff}) yields $E_{\rm eff}(\lambda_{\rm min})$ for
the linear potential. In the ultrarelativistic limit $m=0$ we find in this
way the (variational) effective energy eigenvalues
$$
E_{\rm eff}(\lambda_{\rm min})
= 3\left(\ell + \mbox{$\frac{3}{2}$}\right)^{1/4}
\sqrt{\frac{\Gamma(\ell + 2)}
{\Gamma\left(\ell + \mbox{$\frac{3}{2}$}\right)}\,a}\ .
$$

Table \ref{tab:ratio-linear} compares the nonrelativistic and effectively
semi-relativistic approaches for the linear potential, confirming thereby
again the above findings 1 and 2.
{\normalsize
\begin{table}[hbt]
\begin{center}
\caption[]{Ratio $R$ of the differences between effective and
semi-relativistic and between nonrelativistic and semi-relativistic energy
eigenvalues, defined in Eq.~(\ref{eq:rodoel}), for the three lowest-lying
energy eigenstates (denoted by 1S, 1P, and 2S) of the linear potential $V(r)
= a\,r$, with the slope $a = 0.211 \mbox{ GeV}^2$ and increasing mass $m$ of
the bound-state constituents.}\label{tab:ratio-linear}
\vspace{0.5cm}
\begin{tabular}{|c|c|c|c|}
\hline\hline
&\multicolumn{3}{c|}{}\\[-1ex]
&\multicolumn{3}{c|}{State}\\
&\multicolumn{3}{c|}{}\\[-1.5ex]
\cline{2-4}
$\begin{array}{c}\mbox{$\qquad m$ [GeV]$\qquad$}\\[1ex]{}\end{array}$
&1S&1P&2S\\[-0.5ex]
\hline
&&&\\[-1.1ex]
$\qquad0.250\qquad$&$\qquad0.603\qquad$&$\qquad0.452\qquad$&
$\qquad0.466\qquad$\\
$\qquad0.336\qquad$&$\qquad0.750\qquad$&$\qquad0.576\qquad$&
$\qquad0.572\qquad$\\
$\qquad0.500\qquad$&$\qquad1.013\qquad$&$\qquad0.802\qquad$&
$\qquad0.750\qquad$\\
$\qquad0.750\qquad$&$\qquad1.411\qquad$&$\qquad1.144\qquad$&
$\qquad1.002\qquad$\\
$\qquad1.000\qquad$&$\qquad1.825\qquad$&$\qquad1.498\qquad$&
$\qquad1.253\qquad$\\
$\qquad1.800\qquad$&$\qquad3.304\qquad$&$\qquad2.757\qquad$&
$\qquad2.117\qquad$\\[1.3ex]
\hline\hline
\end{tabular}
\end{center}
\end{table}}

Inspecting once again the large-$\ell$ behaviour of the predicted energy
eigenvalues, we may read off from the above explicit expression for the
ultrarelativistic (variational) effective energy eigenvalues
$E_{\rm eff}(\lambda_{\rm min})$, with the help of a useful relation
describing the asymptotic behaviour of the ratio of gamma functions
\cite{abramow}, viz.,
$$
\lim_{\ell \to \infty}\frac{\Gamma(\ell + z)}{\Gamma(\ell + u)}
= \ell^{z - u}\ ,
$$
the very pleasing result
$$
E_{\rm eff}^2(\lambda_{\rm min}) = 9\,a\,\ell\ .
$$
Accordingly, the effectively semi-relativistic energy eigenvalues of the
linear potential are perfectly able to reproduce the observed linearity of
the Regge trajectories with, however, a slope which is slightly larger than
the one obtained within different investigations \cite{kang75,lucha91regge}
based on the proper semi-relativistic Hamiltonian (\ref{eq:hsemi-rel}), all
of which end up with one and the same finding:
$$
E^2 = 8\,a\,\ell\ .
$$
Moreover, from the point of view of a correct description of the linear Regge
trajectories, both of the semi-relativistic treatments are clearly superior
to the corresponding nonrelativistic approach, which gives for the energy
eigenvalues of the linear potential \cite{lucha91,lucha92}
$$
E_{\rm NR} =
3\left(\frac{a^2}{4\,m}\right)^{1/3}\ell^{2/3} + \mbox{const.}
$$
and therefore
$$
E_{\rm NR}^2(\ell) \propto \ell^{4/3}\ .
$$

\subsection{Funnel potential}

Unfortunately, the potentials considered up to now are merely of more or less
academic interest. Finally, however, we would like to discuss a potential
which has been among the first ones to be proposed \cite{eichten75} for the
description of hadrons as bound states of constituent quarks, viz., the
funnel (or Cornell or Coulomb-plus-linear) potential.

This funnel potential comprehends the two basic ingredients of any realistic,
that is, phenomenologically acceptable, inter-quark potential, namely,
\begin{itemize}
\item at ``short'' inter-quark distances some Coulomb-like singularity of
perturbative origin, which arises from one-gluon exchange, and
\item at ``large'' inter-quark distances an (approximately) linear rise of
non-perturbative origin, which is obviously responsible for colour
confinement.
\end{itemize}
The funnel potential incorporates these two features in the simplest
conceivable manner:
$$
V(r) = - \frac{\kappa}{r} + a\,r\ .
$$
In this form it still represents the prototype of almost all forthcoming
potential models designed to describe all the (binding) forces acting between
quarks.\footnote{\normalsize\ For a brief survey see, for instance,
Ref.~\cite{lucha91}.}

This funnel potential is, beyond doubt, not of the power-law type.
Consequently, it cannot be subjected to the general effective formalism
developed so far but deserves a special treatment, which might consist of
some purely numerical approach.

However, as before, we first want to obtain some insight by applying the
variational procedure described in Subsect.~\ref{subsec:varmeth}. The value
$\lambda_{\rm min}$ of the variational parameter $\lambda$ which minimizes
for the case of the above funnel potential the expectation value of the
effective Hamiltonian $\tilde H_{\rm eff}$ with respect to our Gaussian
trial states is (because of the presence of $\tilde m$ only implicitly)
determined by the relation
$$
\lambda_{\rm min}^3 = \frac{\tilde m}{2}\,
\frac{\Gamma(\ell + 1)}{\Gamma\left(\ell + \mbox{$\frac{5}{2}$}\right)}
\left[\kappa\,\lambda_{\rm min}^2 + a\,(\ell + 1)\right]\ .
$$
In the ultrarelativistic limit $m = 0$ this relation fixes
$\lambda_{\rm min}$ to
$$
\lambda_{\rm min}
= \sqrt{\frac{a\,\Gamma(\ell + 2)}{4\sqrt{\ell + \mbox{$\frac{3}{2}$}}\,
\Gamma\left(\ell + \mbox{$\frac{3}{2}$}\right) - \kappa\,\Gamma(\ell + 1)}}
\ ,
$$
which, in turn, implies for the (variational) effective energy eigenvalues of
the funnel potential
$$
E_{\rm eff}(\lambda_{\rm min}) = 2\,\lambda_{\rm min}
\left(3\sqrt{\ell + \mbox{$\frac{3}{2}$}} - \kappa\,\frac{\Gamma(\ell + 1)}
{\Gamma\left(\ell + \mbox{$\frac{3}{2}$}\right)}\right)\ .
$$

Table \ref{tab:ratio-funnel} compares the nonrelativistic and effectively
semi-relativistic approaches for the funnel potential, illustrating thereby
once more the above findings 1 and 2.
{\normalsize
\begin{table}[hbt]
\begin{center}
\caption[]{Ratio $R$ of the differences between effective and
semi-relativistic and between nonrelativistic and semi-relativistic energy
eigenvalues, defined in Eq.~(\ref{eq:rodoel}), for the three lowest-lying
energy eigenstates (denoted by 1S, 1P, and 2S) of the funnel potential $V(r)
= - \kappa/r + a\,r $, with Coulomb coupling constant $\kappa = 0.456$, slope
$a = 0.211 \mbox{ GeV}^2$, and increasing mass $m$ of the bound-state
constituents.}\label{tab:ratio-funnel}
\vspace{0.5cm}
\begin{tabular}{|c|c|c|c|}
\hline\hline
&\multicolumn{3}{c|}{}\\[-1ex]
&\multicolumn{3}{c|}{State}\\
&\multicolumn{3}{c|}{}\\[-1.5ex]
\cline{2-4}
$\begin{array}{c}\mbox{$\qquad m$ [GeV]$\qquad$}\\[1ex]{}\end{array}$
&1S&1P&2S\\[-0.5ex]
\hline
&&&\\[-1.1ex]
$\qquad0.250\qquad$&$\qquad0.629\qquad$&$\qquad0.463\qquad$&
$\qquad0.500\qquad$\\
$\qquad0.336\qquad$&$\qquad0.764\qquad$&$\qquad0.584\qquad$&
$\qquad0.603\qquad$\\
$\qquad0.500\qquad$&$\qquad0.988\qquad$&$\qquad0.801\qquad$&
$\qquad0.771\qquad$\\
$\qquad0.750\qquad$&$\qquad1.288\qquad$&$\qquad1.118\qquad$&
$\qquad0.992\qquad$\\
$\qquad1.000\qquad$&$\qquad1.559\qquad$&$\qquad1.437\qquad$&
$\qquad1.191\qquad$\\
$\qquad1.800\qquad$&$\qquad2.277\qquad$&$\qquad2.489\qquad$&
$\qquad1.770\qquad$\\[1.3ex]
\hline\hline
\end{tabular}
\end{center}
\end{table}}

The large-$\ell$ behaviour of the ultrarelativistic (variational) effective
energy eigenvalues $E_{\rm eff}(\lambda_{\rm min})$ resulting from this
expression is the same as for the pure linear potential:
$$
E_{\rm eff}^2(\lambda_{\rm min}) = 9\,a\,\ell\ .
$$
This circumstance is an unavoidable consequence of the fact that in the limit
$\ell \to \infty$ all contributions of the Coulomb part of the funnel
potential to the above effective energy eigenvalues $E_{\rm eff}$ vanish,
which may be seen immediately by recalling once more the above-mentioned
asymptotic behaviour of the ratio of gamma functions. Accordingly, for large
orbital angular momenta $\ell$ the positioning of the energy levels of the
funnel potential is controlled by its confinement part
only.\footnote{\normalsize A similar observation has already been made in
Ref.~\cite{lucha91regge} within a slightly different context
\cite{lucha91rel,lucha90com}.}

\section{Conclusions}\label{sec:effhamconcl}

The present paper has been dedicated to the formulation of effectively
semi-relativistic Hamiltonians which are designed in such a way that---by a
suitable interpretation of their (effective) parameters---they allow us to
approximate an entirely semi-relativistic formalism at a formally
nonrelativistic level. Application of the developed formalism to a few
representative static interaction potentials gave indications that below some
specific critical mass of the involved particles, where our effective energy
eigenvalues are closer to the (exact) semi-relativistic ones than those of a
nonrelativistic description, the effective approach represents, at least in
relativistic regions, an improvement of the certainly rather crude
nonrelativistic approximation. Simultaneously, this observation might
contribute to the eventual explanation of the surprising success of (a
variety of) nonrelativistic potential models in describing hadrons as bound
states of quarks even for the case of relativistically moving constituents.

\end{document}